\input{aipcheck}

\documentclass[
    ,final            
  ]
  {aipproc}

\layoutstyle{8x11single}

\begin{document}

\title{Ion-Neutral Collisions  in the Interstellar Medium: Wave Damping and Elimination of Collisionless Processes}


\author{Steven R. Spangler}{
  address={Department of Physics and Astronomy, University of Iowa}
}

\author{Allison H. Savage}{
  address={Department of Physics and Astronomy, University of Iowa}
}

\author{Seth Redfield}{
  address={Department of Astronomy and Van Vleck Observatory, Wesleyan University}
  }

\begin{abstract}
Most phases of the interstellar medium contain neutral atoms in addition to ions and electrons.  This introduces differences in plasma physics processes in those media relative to the solar corona and the solar wind at a heliocentric distance of 1 astronomical unit.  In this paper, we consider two well-diagnosed, partially-ionized interstellar plasmas.  The first is the Diffuse Ionized Gas (DIG) which is probably the extensive phase in terms of volume.  The second is the gas that makes up the Local Clouds of the Very Local Interstellar Medium (VLISM).  Ion-neutral interactions seem to be important in both media.  In the DIG, ion-neutral collisions are relatively rare, but sufficiently frequent to damp magnetohydrodynamic (MHD) waves (as well as propagating MHD eddies) within less than a parsec of the site of generation.  This result raises interesting questions about the sources of turbulence in the DIG.  In the case of the VLISM, the ion-neutral collision frequency is higher than that in the DIG, because the hydrogen is partially neutral rather than fully ionized.  We present results showing that prominent features of coronal and solar wind turbulence seem to be absent in VLISM turbulence.  For example, ion temperature does not depend on ion mass.  This difference may be attributable to ion-neutral collisions, which distribute power from more effectively heated massive ions such as iron to other ion species and neutral atoms.  
\end{abstract}

\maketitle


\section{INTRODUCTION}
The interstellar medium (ISM) consists of several ``phases'' of matter at different temperatures and densities.  In most or all of these phases, the gas is sufficiently ionized to be considered a plasma.  A difference with the well-studied media of the solar wind and the solar corona is that many of the phases, and specifically the two we will discuss here, are partially ionized.  In addition to one or more ion species and electrons, there are neutral atoms and associated ion-neutral collisions.  These ion-neutral collisions cause processes which are not present in the fully ionized corona and solar wind at a heliocentric distance of one astronomical unit \footnote{Neutral atoms from interstellar space are present in the solar wind at 1 AU as well, but they do not play an important role there.}. 

This paper will deal with two topics involving the interaction of ions with neutrals in the interstellar medium.  
\begin{enumerate}
\item Ion-neutral collisions constitute a damping mechanism for magnetohydrodynamic (MHD) waves and turbulence in the interstellar medium.  As a result, turbulence propagating from a central source of turbulence, such as a supernova remnant, is limited in how far it can go before being attenuated.  In the case of the Diffuse Ionized Gas (DIG) component of the ISM \citep{Cox87,Haffner03}, which is probably the most extensive in terms of volume and contains the best-diagnosed interstellar turbulence, the dissipation lengths are short and place interesting constraints on the sources and mechanisms for interstellar turbulence. Alternatively, this damping length calculation could point to odd features of turbulent dispersal in the interstellar medium.  This issue was discussed four years ago \cite{Spangler07}.  It remains an interesting topic worthy of attention, and there have been some developments in the last four years which are relevant to this topic.  
\item The nearest parts of the interstellar medium are the clouds of the Very Local Interstellar Medium (VLISM) \citep{Frisch00,Redfield09}. There are 15 of these clouds within 15 parsecs of the Sun \citep{Redfield08}, and they have typical dimensions of a parsec to several parsecs.  The reason for discussing them in the present context is that astronomical spectroscopic observations  \citep[e.g.][]{Redfield01} have provided very good plasma diagnostics, including information on turbulence in the clouds.  An obvious exercise is to compare the turbulence in these clouds with the paradigmatic turbulence in the solar corona and solar wind.  As we will see, there may be important differences between heliospheric turbulence and turbulence in the Local Clouds.  The strongest result is that signatures of collisionless plasma processes in the solar wind and corona are absent in the Local Clouds.   As we discuss below, it is plausible that ion-neutral collisional processes are responsible for this difference.  \end{enumerate}
\subsection{The microphysics of ion-neutral collisions}
The subject of plasma physics may be defined as the dynamics of ions and electrons.  Ions and electrons generate electrical currents and charge densities and respond to the electric and magnetic fields in plasma waves.  Neutral atoms and molecules do neither of these things, and interact with the plasma only via collisions.  An important role of neutrals, discussed in this paper, is to constitute a stationary or independently-moving background with which ions (and electrons) collide.  These collisions transfer energy from the directed motion of plasma waves or turbulence into disordered thermal motion.  These collisions thus constitute a damping mechanism for the plasma waves and turbulence.  There are two types of ion-neutral collisions which will concern us here. In both cases, a quantity of interest is the collision frequency 
\begin{equation}
\nu_0 = \sigma(v) v n
\end{equation}
where $v$ is the ion speed, $\sigma(v)$ is the velocity-dependent cross section and $n$ is the number density of target atoms. In some applications, an average collision frequency may be taken by averaging the speed-dependent collision frequency over the ion distribution function. 
\paragraph{Charge Exchange} In the process of charge exchange, an ion strips the electron from a passing neutral atom, so the ion and neutral atoms change roles. This mechanism can and generally does (in the space science context) occur with negligible change in the energy or momentum of the nuclei.  In this respect it differs from Coulomb scattering, in which at least one of the colliding ions undergoes a substantial change in momentum.  The expression for the charge exchange cross section is 
\begin{equation}
\sigma_{CX} = 2 \pi \int_0^{\infty}R P(R) dR
\end{equation}
where $R$ is the impact parameter between the colliding atom and ion, and $P(R)$ is the quantum mechanical probability for the electron transfer at distance $R$.  For protons on light neutral atoms such as H and He, and for atomic speeds characteristic of the ISM, 
\begin{equation}
\sigma_{CX} \simeq (3-6) \times 10^{-15} \mbox{ cm}^2
\end{equation}
depending on the species interacting and the speed of interaction. 
\paragraph{Induced Dipole Scattering} An ion approaching a neutral atom or molecule causes a polarization of the charge cloud within the atom, leading to a dipole moment, even if the unperturbed atom or molecule possesses no electric dipole moment.  The dipole electric field of the atom or molecule then affects the motion of the ion, leading to scattering.  The strength of the interaction depends on the polarizability $\alpha$ of the neutral atom.  The cross section for this process is 
\begin{equation}
\sigma_{ID}(v) = \frac{2.21 \pi}{v} \sqrt \frac{\alpha e^2(M+m)}{Mm}
\end{equation}
where $m$ and $M$ are the masses of the ion and the  neutral, respectively \citep{Banks66}. For hydrogen ions scattering from neutral hydrogen or helium, the cross section is  $\sigma_{ID}(v) \simeq (1-2) \times 10^{-15} \mbox{ cm}^2$ for typical ISM thermal speeds. The range depends on the target neutral (scattering from H$^0$ has a larger cross section) and the precise value of the speed of interaction.  Interestingly, charge exchange and induced dipole collisions have similar values for the cross section for interstellar conditions.  For the media of interest in this paper, the charge exchange cross section is slightly larger, but the values depend on temperature.  For lower-temperature parts of the interstellar medium, induced dipole scattering is more significant \citep{Banks66}.  

A possibly significant difference between the two types of interaction is that charge exchange does not produce any true randomization of atomic motion, only an exchange of the roles of neutral and ion.  Induced dipole scattering, on the other hand, does produce a randomization of initially ordered motion.  We now discuss two topics in the physics of the interstellar medium in which ion-neutral collisions play an important role. 
  
\section{THE PROPAGATION DISTANCE OF INTERSTELLAR TURBULENCE} We first consider the sources of turbulence in the Diffuse Ionized Gas (DIG) component of the interstellar medium.  The DIG probably occupies the largest fraction of the volume in the interstellar medium \citep{Cox87,Haffner03} and is almost certainly the host plasma for the turbulent density fluctuations which produce radio scintillations of pulsars and extragalactic radio sources.  These radio scintillations provide our best information on the nature of interstellar turbulence \citep{Rickett90,Armstrong95}.  The basic plasma properties of the DIG are given in Table 1 \citep{Minter97b}.  One of the most important properties in Table 1 is the helium ionization fraction; astronomical spectroscopic observations indicate that helium is only partially ionized \citep{Reynolds95,Madsen06}.  It may be completely neutral, but is probably not more than 50 \% ionized. In fact, a best estimate would seem to be that the helium in the DIG is $\leq 30$ \% ionized \citep{Madsen06}.  Hydrogen ions, moving as part of Alfv\'{e}nic waves or turbulence, collide with these neutral helium atoms and convert directed motion into random motion.  

\begin{table}
\begin{tabular}{lr}
\hline
\tablehead{1}{r}{b}{Plasma characteristic}
  & \tablehead{1}{r}{b}{Value}   \\
\hline
plasma density & 0.08 cm$^{-3}$\\
magnetic field strength & 3-5 $\mu$ Gauss\\
temperature & 8000 K\\
Alfv\'{e}n speed & 23 km/sec \\
Helium ionization fraction & 0 - 50 \%\\
\hline
\end{tabular}
\caption{Plasma properties of the Diffuse Ionized Gas (DIG)}
\end{table}
Previous investigations have indicated that ion-neutral collision processes might play an important thermodynamic role in the DIG.  The estimated volumetric heating rate due to ion-neutral collisions is plausibly comparable to the radiative cooling rate of this medium \citep{Spangler91,Minter97b,Minter97a}. 

If, as has been argued by the above authors, ion-neutral collisional damping of turbulence is an important process in the DIG, it immediately raises an interesting question regarding the source or sources of this interstellar turbulence.  What astronomical objects are the sources of interstellar turbulence, and how does this turbulence propagate or diffuse to fill the interstellar medium?  This last statement is motivated by radio scattering observations showing that the turbulence responsible for radio scintillations seems to fill the DIG plasma.  If the DIG plasma is heated by dissipating turbulence, this process of dissipation also limits the propagation distance of the turbulence from the sources of turbulence.  In what follows, we assume that the DIG turbulence consists of either Alfv\'{e}n waves or turbulent eddies that propagate along the magnetic field at the Alfv\'{e}n speed $V_A$.  

The propagation distance $l$ (or equivalently, dissipation length) of turbulence is 
\begin{equation}
l = \frac{V_A}{\gamma}
\end{equation}
where $\gamma$ is the damping rate of the MHD wave or eddy.  We adopt the damping rate for Alfv\'{e}n waves given by \cite{Kulsrud69}, which is 
\begin{equation}
\gamma = \frac{\nu_0}{2}
\end{equation}
where $\nu_0$ is the collision frequency defined in Equation (1).  Equation (6) is valid only for Alfv\'{e}n waves with frequencies $\nu \gg \nu_0$, or wavelengths much smaller than the wavelength of an  Alfv\'{e}n wave with frequency equal to $\nu_0$.  We use the calculation for $\nu_0$ in \cite{Spangler91} to obtain $\nu_0 = 6.5 \times 10^{-12}$ Hz.  Using this value, and an estimate of the Alfv\'{e}n speed of 23 km/sec \citep[][also see Table 1]{Minter97b}, we get a propagation distance of $l \simeq 3.5 \times 10^{17} \mbox{ cm} \simeq 0.1$ parsecs. 

The previous calculation used the value for the collision frequency $\nu_0$ taken from \cite{Spangler91}, based on the cross section for the induced dipole scattering process.  As noted in Section 1, the cross section for charge exchange is a factor of a few larger for DIG conditions, and would thus decrease the propagation distance $l$ by a similar factor.  In any case, the small propagation distance can be compared with the much larger distances to a typical supernova remnant. Supernova remants are obvious candidate objects for generation of interstellar turbulence because they are sites of large energy input to the interstellar medium and are subject to a number of hydrodynamic and magnetohydrodynamic instabilities.  Blair et al \citep{Blair99} report that the Cygnus Loop, perhaps the closest young, supersonically-expanding supernova remnant, is 440 parsecs away.  Obviously, turbulence generated by the Cygnus Loop could not propagate through a homogeneous DIG to the location of the Sun.  
\subsection{Possible resolutions of the short propagation length problem}
The arguments above  were given by \cite{Spangler07}, who also discussed  possible, qualitative processes which could evade these problems. These suggestions were illustrated in cartoon form in Figure 2 of that paper. 
\begin{enumerate}
\item There may be unknown local turbulence generation mechanisms spread widely through the interstellar medium.  For example, vortex sheets may be widespread through the ISM, and the associated velocity shear might generate turbulence.  
\item The above calculation assumes a uniform interstellar medium with the same neutral density everywhere.  It is conceivable that that the extreme inhomogeneity which characterizes all aspects of the interstellar medium also applies to the ionization fraction.  In this case, there might be channels or conduits of completely ionized plasma, and turbulence might propagate throughout the ISM via these channels.  This possibility was labeled ``wave percolation'' in analogy with the transport of water through highly inhomogeneous, fractured rock \citep{Spangler07}.  
\end{enumerate}

\subsection{Developments in the last four years} The discussion in \cite{Spangler07} was prepared for and presented at a meeting in Socorro, New Mexico in 2006, dealing with pronounced inhomogeneity in the ISM. Since that time, there have been developments in observational astronomy which are relevant to this discussion.  
\begin{enumerate}
\item An extremely significant development in the field of interstellar scintillations is the realization that much, and conceivably all, of pulsar radio wave scintillation is produced by density fluctuations in highly localized, turbulent sheets, rather than fluctuations that are uniformly distributed throughout the interstellar medium.  This major evolution in our view of the turbulence responsible for scintillations came about as a result of the discovery of ``pulsar arcs'' in the secondary dynamic spectra of pulsars.  This phenomenon was discovered by Daniel Stinebring and his students at Oberlin College; expositions showing the ubiquity, if not universality of the phenomenon as well as a theoretical explanation of the arcs is given in \cite{Walker04} and \cite{Cordes06}.  To state matters briefly,  the physics of the arcs seems to require that the density fluctuations responsible for the scintillations be confined to a sheet extending over a small fraction of the line of site. Most pulsars studied to date seem to have one, or at most a few arcs in their secondary spectra, so it seems that interstellar turbulence is concentrated in a few intense, highly localized regions.  This observation would seem to be qualitatively consistent with the notion that there are spatially-intermittent generators of turbulence, rather than propagation of turbulence throughout the ISM from a relatively small number of sources in the Galaxy. 
\item Linsky et al \cite{Linsky08} contend that the turbulence responsible for rapid scintillations of the quasars B1257-326, B1519-273, and J1819+385 is contained in a region of interaction between the LIC cloud and G cloud in the Very Local Interstellar Medium (VLISM, more discussion below).  This would also support the idea that turbulence in the ISM is restricted to special, highly localized regions.  
\end{enumerate} 
Both of these recent developments support the first suggestion above, i.e. that regions of intense turbulence are generated by local conditions far from the ``energy drivers'' of the interstellar medium.  
\section{A COMPARISON OF SOLAR WIND AND VERY LOCAL INTERSTELLAR TURBULENCE} The vicinity of the Sun is dominated by a ``Local Cavity'' in the interstellar medium, characterized by much lower than average densities \citep{Lallement03,Welsh10}, and perhaps higher temperatures.  However, in the immediate vicinity of the Sun there are many small, tenuous clouds \citep{Frisch00,Redfield09}.  These clouds do not seem to extend beyond the vicinity of the Sun. Redfield and Linsky \citep{Redfield08} report 15 of these clouds within 15 parsecs of the Sun.  There are a number of interesting aspects of these clouds, such as the fact that the Sun is within, though close to the edge of one of them, the Local Interstellar Cloud (LIC) \citep{Redfield08}.  

Observations that show the existence of these clouds and provide diagnostics for their properties come from very high resolution astronomical spectroscopy \citep{Redfield01}.  The spectrometers used in this study, primarily the STIS instrument on the Hubble Space Telescope, have resolving powers of $R \equiv \frac{\lambda}{\Delta \lambda} \simeq 100,000$.  This allows absorption lines in the spectra of nearby stars to be resolved.  Many cases have been found of absorption lines in the spectra of nearby stars which are produced by clouds between us and these stars.  For example, Figure 1 shows the spectrum of such a star, HD29419, showing absorption lines corresponding to two local clouds \citep{Redfield01}.  
\begin{figure}
  \includegraphics[height=.3\textheight]{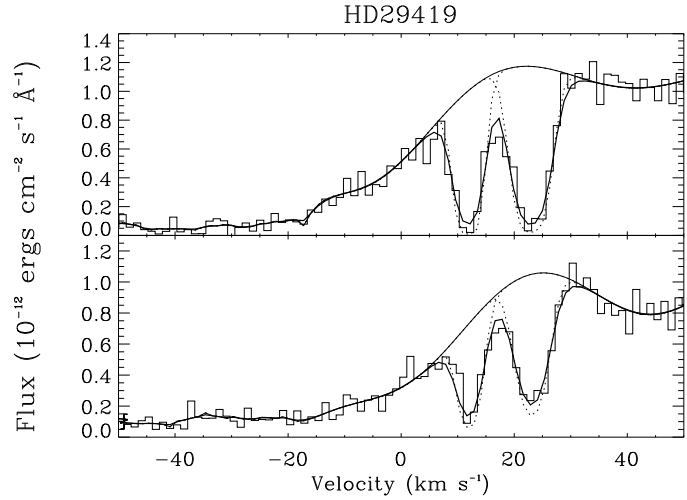}
  \caption{Absorption spectrum of the star HD29419 in the h and k lines of MgII. There are two absorption lines in its spectrum, produced by two local clouds. Excerpted from \cite{Redfield01}. This version of the figure is taken from \cite{Spangler10}.}
\end{figure}
In this figure, the thin solid line gives an estimate of the stellar photospheric continuum.  There are two absorption components, one corresponding to the Local Interstellar Cloud (LIC), and the other the Hyades Cloud.  For each absorption component, a fit gives the line centroid, corresponding to the Doppler velocity of the relevant cloud, the strength of the line, determined by the column density of the atom or ion producing the line, and finally the Doppler width of the line. It is this Doppler width $b$ which is of primary interest to us. Diagnostics provided by these lines, as well as in-situ measurements of neutral atoms which flow into the inner solar system, have allowed remarkably good specification of basic plasma parameters in these clouds.  These properties are summarized in Table 2. 
\begin{table}
\begin{tabular}{lr}
\hline
\tablehead{1}{r}{b}{Plasma characteristic}
  & \tablehead{1}{r}{b}{Value}   \\
\hline
plasma density & 0.11 cm$^{-3}$\\
neutral density & 0.1 cm$^{-3}$\\
magnetic field strength & 3-5 $\mu$ Gauss (assumed)\\
temperature & 4000 - 8000 K\\
proton gyrofrequency & $(3-5) \times 10^{-2}$ Hz  \\
ion-neutral collision frequency & $3 \times 10^{-10}$ Hz \\
collisional mean free path & $5 \times 10^{15}$ cm = $1.5 \times 10^{-3}$ parsec \\
\hline
\end{tabular}
\caption{Plasma properties of the Local Clouds}
\end{table} 
\subsection{Inferring cloud turbulence properties from high-resolution spectroscopy} 
Our information about turbulence in the Local Clouds comes almost exclusively from measurements of the absorption line widths $b$.  It is remarkable that we can tell anything about turbulence from such measurements.  We can do so because the same absorption component can be measured in lines of up to 8 ions and atoms.  Redfield and Linsky  \citep{Redfield04} model the measured line profile for all species by the expression
\begin{equation}
b^2 = \frac{2k_B T}{m} + \xi^2
\end{equation} 
where $m$ is the mass of an atom or ion, $T$ is the temperature, and $\xi$ is the nonthermal component of the Doppler width due to random motions of the gas.  We identify $\xi$ with the rms velocity fluctuation due to turbulent fluctuations of all spatial scales. With measurements of the line width for more than 2 atoms or ions, we can fit for $T$ and $\xi$.  With 8 transitions, one has a very good estimate of the significance and goodness of the fit.  

Equation (7) might strike a solar astronomer or space physicist as naive.  In the solar corona and, to a lesser extent the solar wind, there is not a single temperature which characterizes all ion species.  Observations in the solar corona show that more massive ions have higher temperatures than less massive ones \citep{Cranmer02}. The physically relevant quantity is the ion cyclotron frequency, or the ion inertial scale corresponding to a certain ion.  As a prominent example, at heliocentric distances of 2 - 3 $R_{\odot}$ the proton temperature is 1 - 2 million K, whereas that of OVI is of order 100 million K \citep{Cranmer02}.  The ratio of  temperatures is more than proportional to the ion mass ratio.  In addition, the heating is largely perpendicular to the large scale magnetic field.  Observations in the corona indicate that $T_{\perp} \gg T_{\parallel}$.  In-situ observations in the solar wind show similar, though less pronounced phenomena.  There is evidence for enhanced energy input to heavier ions, and the perpendicular temperature can exceed the parallel temperature, or at least be higher than would be expected, given adiabatic cooling in the expansion of the solar wind. 
The explanation generally given for this variety of temperatures is ion-cyclotron resonance interaction between turbulence and charged particles \citep{Hollweg08,Cranmer02}.  This is discussed further below.  

The argument for use of a single temperature in the Local Clouds is that Equation (7) provides a very good, and statistically satisfactory, fit to the data. For example, Figure 2 shows the line width data for the absorption component in the spectrum of the star Capella.  
\begin{figure}
  \includegraphics[height=.3\textheight]{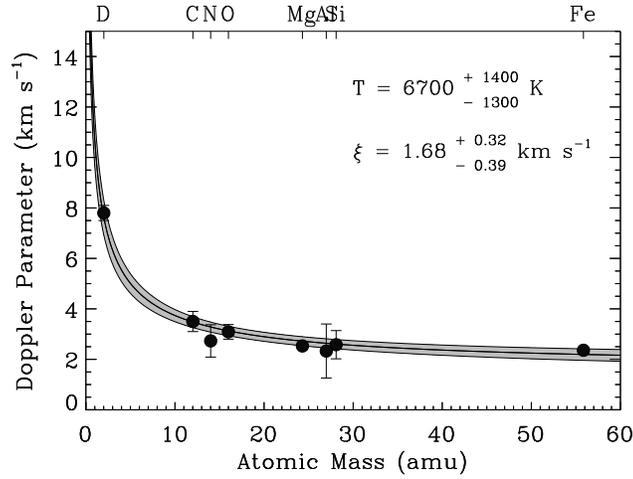}
  \caption{Line width parameter $b$ as a function of atomic mass for the line of sight to the star Capella.  The solid curve gives the least-squares fit of Equation (7) to the data, and the shaded region indicates the 1$\sigma$ range of uncertainty. Excerpted from \cite{Redfield04}. This version of the figure is taken from \cite{Spangler10}.}
\end{figure}
This figure provides no motivation for considering a more complex temperature model to describe the data.  Equation (7) is entirely adequate in a least-squares sense.  Results similar to those shown in Figure 2 are available for many more stars; see Figure 1 of \cite{Redfield04}.  Furthermore, in addition to being an entirely adequate representation of the data, these data actually preclude, at some level, a temperature which depends on ion mass.  We are currently carrying out an analysis to determine the maximum allowable degree of mass-dependence of the temperature compatible with the data. We consider a generalization of Equation (7) which permits the temperature to depend on ion or atomic mass,
\begin{equation}
b^2 = \frac{2k_B T}{m} \left( \frac{m}{m_0} \right)^d + \xi^2
\end{equation} 
where $m_0$ is the mass of a fiducial ion or atom (say deuterium) and $d$ is a parameter which describes the degree of mass-dependence of the temperature.  Obviously, Equation (8) contains Equation (7) as a special case with $d=0$.  Mass-proportional ion temperature corresponds to $d=1$.  For the case of the hydrogen and OVI temperatures in the solar corona, $d > 1$.  In our analysis of Local Cloud data, we are restricting consideration to those lines of sight which have measurements for the full set of 8 lines. For each line of sight, we determine the maximum value of $d$ which is statistically-compatible with the data in a least-squares sense.  The parameters $T$ and $\xi$ are left relatively unconstrained.  These maximum allowable values of $d$ range from about 0.3 to 0.8, depending on the line of sight.  Fits with $d > 0$ have lower values of $\xi$, often significantly so.  In no case does Equation (8) provide a satisfactory fit when Equation (7) could not.  To summarize these preliminary results, a single temperature $T$ for all species is a satisfactory representation of the data, and a temperature proportional to ion mass appears to be ruled out in all cases.     

An even more restrictive result which emerges from the fits presented in \cite{Redfield04} is that the temperature is the same for both ions and neutral atoms.  In Figure 2 and the more extensive set of fits and data shown in Figure 1 of \cite{Redfield04}, the lines of D, N, and O are for neutral atoms, while the lines of C, Mg, Al, Si, and Fe are of ions.  Obviously, any plasma wave-particle interaction, resonant or non-resonant, will affect only charged particles, and will not directly affect the neutral atoms.  

A final remark pertains to the existence of a temperature anisotropy in the Local Clouds.  As noted above, perpendicular heating, or an enhancement of the perpendicular temperature, appears to be a general characteristic of the solar wind.  If perpendicular heating were occurring in the Local Clouds, and {\em if} there is a uniform direction of the Galactic magnetic field across the system of Local Clouds\footnote{The notion that the Galactic vector magnetic field, consisting of a truly Galactic-scale component plus a turbulence component, would be relatively constant across the system of Local Clouds, i.e. an extent of perhaps 30 parsecs, is a strong assumption but a plausible one.  Its validity is largely dependent on the value of the outer scale in the interstellar turbulence.}, then the temperature $T$ should depend on direction in the sky.  In certain directions we would be looking across the Galactic field and should see predominantly $T_{\perp}$, while for two parts of the sky aligned with the Galactic field we should measure $T_{\parallel}$.  A casual examination of this point is possible via Figure 8 of \cite{Redfield04}.  Although random variations in $T$ are seen from one line of sight to another, there is no evidence in this figure for a systematic anisotropy which would indicate the presence of temperature anisotropy and a uniform Galactic magnetic field. A quantitative study of this point, together with limits to the degree of temperature anisotropy is in progress. 

In summary, the data of \cite{Redfield04} for turbulence and plasma parameters in the Local Clouds suggest that the plasma wave-particle effects which are so prominent in the corona and solar wind are absent in the Local Clouds.  In the next section, we briefly comment on possible explanations of this. 
\subsection{Reasons for the absence of collisionless plasma processes in the Local Clouds} The turbulence in the Local Clouds of the ISM, as diagnosed by spectral line widths, appears to differ in a number of respects from coronal and solar wind turbulence \citep{Spangler10}.  In particular, the characteristics attributed to collisionless processes for the interaction of charged particles and turbulence, such as ion mass dependent temperatures and large perpendicular-to-parallel temperature ratios, are not present. These properties have generally been attributed to ion cyclotron resonance processes in wave-particle interactions \citep{Hollweg08,Cranmer02}. In this section, we briefly consider possible explanations for this.  The simplest answer is probably that the Local Clouds are collisional via the ion-neutral collision processes discussed in Section 1, whereas the solar corona at heliocentric distances $\geq 2.0 R_{\odot}$ and the solar wind at 1 astronomical unit are collisionless plasmas.  

To accept this suggestion, we need to convince ourselves that the partially ionized plasma media of the Local Clouds are collisional.  Following the suggestions of Uzdensky \citep{Uzdensky07}, we can define a collisional plasma in two ways.  
\begin{enumerate}
\item We can first ask if the ion cyclotron frequency greatly exceeds the ion-neutral collision frequency.  If this is the case, processes which involve ion-cyclotron resonances can proceed through many cyclotron periods before being disrupted by a collision.  For example, a distribution function which is unstable to the growth of Alfv\'{e}n or Fast Mode waves will generate the unstable waves, or large amplitude, resonant waves will substantially modify an ion distribution function on time scales shorter than the collision time. According to this way of defining things, a plasma in which an ion cyclotron frequency is much higher than the ion-neutral collision frequency would be collisionless for that ion.  
\item The second criterion of collisionality according to Uzdensky considers the mean free path for collisions and the size of the plasma.  If the mean free path for collisions is much larger than the size of the cloud, the typical ion in the medium would not have undergone a collision, and one would expect collisionless physics to be applicable. In the opposite limit of the mean free path being much smaller than the size of the cloud, the typical ion would have undergone numerous collisions.  It seems plausible that collisions would redistribute the energy gained from collisionless processes among different parts of the distribution function, and indeed different species in the plasma. Support for these ideas may be found in \cite{Kasper09}, where it is noted that collisional parts of the solar wind at 1 AU, such as dense parts of the interplanetary current sheet, lack the collisionless indicators of temperature anisotropy and mass-dependent temperature.   
\end{enumerate}
We have enough information about the Local Clouds to apply the above criteria. In Table 2 we list the proton ion cyclotron frequency, ion-hydrogen collision frequency, and mean free path, which are calculated from the plasma parameters listed previously in the table.  In carrying out the calculations leading to these numbers, we assume a collision cross section of $2 \times 10^{-15} \mbox{cm}^{-3}$, which is appropriate for  induced dipole scattering of H and H$^+$ at a relative speed of 14 km/sec.  If we assume that the most important ion-neutral process is charge exchange, the cross section would be a few times larger, resulting in  a collisional mean free path a few times smaller. Table 2 shows that the Local Clouds strongly fulfill the criterion for a collisionless plasma according to the first definition, and strongly fail it according to the second.  

Given the above considerations, we can say that the absence in the Local Clouds of those signatures of collisionless plasma dynamics which are seen in the solar corona and solar wind is plausibly due to ion-neutral collisions which redistribute the energy placed in ion distribution functions by plasma waves and turbulence. While this might seem like a trivial result according to criterion \# 2, our investigation indicates that there are not vigorous ion cyclotron resonance processes occurring at small scales in the Local Clouds.   

In concluding this section, it is interesting to note a recent result which constitutes an alternative description of the same data.  Chandran  \cite{Chandran10} has  proposed that the features of coronal and solar wind turbulent heating that we have nominated ``collisionless'' arise not through an ion cyclotron resonant interaction with turbulence, but rather via stochastic acceleration.  Stochastic acceleration occurs when the turbulent velocity fluctuations $\delta v_i$ on the spatial scale of the ion cyclotron radius become comparable to the perpendicular component of the ion thermal velocity $v_{\perp i}$.  Chandran  \cite{Chandran10} defines a parameter $\epsilon_i$ which is the ratio of these two speeds, 
\begin{equation}
\epsilon_i \equiv \frac{\delta v_i}{v_{\perp i}}
\end{equation}
and contends that stochastic acceleration occurs when $\epsilon \simeq 1$.  He presents arguments that stochastic acceleration, once commenced, will self-regulate at  $\epsilon \simeq 1$.  He also shows that this mechanism can reproduce the observed features which hithero have been attributed to ion cyclotron resonant heating.  

It can be argued that if the mechanism of \cite{Chandran10} is in fact operative in the solar corona and solar wind, it does so because of the small  inertial subrange in the solar corona and solar wind.  It is plausible to assume that the rms velocity fluctuation at the outer scale, $\delta v_0$ is limited to a fraction of the Alfv\'{e}n speed, 
$\delta v_0=b V_A$, where $b \simeq 0.1 - 0.5$.  This is certainly the case in the solar wind.  The amplitude of the  fluctuations decreases with decreasing spatial scale, a fact recognized and utilized in Equation (13) of \cite{Chandran10}. In the case of the solar wind (and presumably in the solar corona) the ratio of the outer scale to the ion inertial length is  2-3 orders of magnitude; a number large enough to allow a true inertial subrange, but small enough so that velocity fluctuations on the ion inertial length might be large enough to permit stochastic acceleration.  

The case in the Local Clouds, or any phase of the interstellar medium, is very different.  Assuming that the outer scale is of the order of a parsec, and the ion inertial length is a few hundred kilometers, the ratio of outer scale to ion inertial scale is of order $10^{10}$, so stochastic acceleration as described in \cite{Chandran10} would not occur.  If one takes this inchoate viewpoint, the absence of phenomena such as temperature anisotropy and mass-proportional temperature in the ISM has nothing to do with ion-neutral interactions, and everything to do with the extent of the intertial subrange in interstellar turbulence.  
\section{DIRECTIONS FOR FUTURE RESEARCH} 
There are a number of interesting research projects which could adumbrate the issues brought up in this paper.  
\begin{enumerate}
\item Additional spectroscopy of stars whose lines of sight pass through the Local Clouds could yield valuable information on the plasma physics of these clouds.  Progress could be made both through measurements of additional transitions of different atoms or ions on the same lines of sight, as well as additional measurements on new lines of sight.  What might be particularly interesting would be comparisons of line broadening on closely-spaced lines of sight, such as to components of a visual binary star.  Data of this sort could indicate the degree of intermittency in Local Cloud turbulence. 
\item We have suggested above that collisions between ions (moving in response to plasma waves) and neutral atoms  can damp the waves and alter the ion distribution functions.  Theoretical research could clarify the resultant changes to ion and neutral distribution functions. Such studies might also reveal observational signatures which could be sought in astronomical observations.  
\item Laboratory experiments should be made to investigate ion-neutral interactions in plasma turbulence.  In the last decade or so, new, larger machines have been made available to plasma experimentalists, and new diagnostic instruments have been developed.  These experimental advances will permit measurements with  machines which are large enough to contain Alfv\'{e}n waves, and possess roughly similar densities of ions and neutrals.  
\item The results of Section 2 indicate that the distribution of plasma turbulence in the interstellar medium might be highly spatially inhomogeneous, and that it is quickly damped when present.  New astronomical observations should be considered which could reveal anomalous neutral heating in the DIG and similar phases of the ISM.  
\end{enumerate}

\begin{theacknowledgments}
 This research was supported at the University of Iowa by grants ATM09-56901 and AST09-07911 from the 
National Science Foundation.
\end{theacknowledgments}


\begin{thebibliography}{9}
\bibitem[Armstrong et al (1995)]{Armstrong95} J.W. Armstrong, B.J. Rickett, and S.R. Spangler, Astrophys. J.~443, 209--221 (1995)
\bibitem[Banks (1966)]{Banks66} P. Banks, Planet Sp. Sci.~14, 1105--1122 (1966)
\bibitem[Blair et al (1999)]{Blair99} W.P. Blair, R. Sankrit, J.C. Raymond, and K.S. Long, Astron
. J.~118, 942--947 (1999)
\bibitem[Chandran (2010)]{Chandran10} B.D.G. Chandran, Astrophys. J.~720, 548--554 (2010)
\bibitem[Cordes et al (2006)]{Cordes06} J.M. Cordes, B.J. Rickett, D.R. Stinebring, and W.A. Coles, Astrophys. J.~637, 346--365 (2006)
\bibitem[Cox and Reynolds (1987)]{Cox87} D.P. Cox and R.J. Reynolds, Annu. Rev. Astr. Ap.~25, 303--344 (1987)
\bibitem[Cranmer (2002)]{Cranmer02} S.R. Cranmer, Space Sci. Rev.~101, 229--293 (2002) 
\bibitem[Frisch (2000)]{Frisch00} P.C. Frisch, Am. Sci.~88, 52--59 (2000) 
\bibitem[Haffner et al (2003)]{Haffner03} L.M. Haffner, R.J. Reynolds, S.L. Tufte, G.J. Madsen, K.P. Jaehnig, and J.W. Percival, Astrophys. J. Supp.~149, 405--422 (2003)
 \bibitem[Hollweg (2008)]{Hollweg08} J.V. Hollweg, J. Astrophys. Astr.~29, 217--237 (2008)
\bibitem[Kasper et al (2009)]{Kasper09} J.C. Kasper, B.A. Maruca, and S.D. Bale, arXiv: 0911.2715 (2009) 
\bibitem[Kulsrud and Pearce (1969)]{Kulsrud69} R.M. Kulsrud and W.P. Pearce, Astrophys. J.~156, 445--469 (1969)
\bibitem[Lallement (2003)]{Lallement03} R. Lallement, B.Y. Welsh, J.L. Vergely, F. Crifo, and D. Sfeir, Astron. \& Astrophys..~411, 447--464 (2003)
\bibitem[Linsky et al (2008)]{Linsky08} J.L. Linsky, B.J. Rickett, and S. Redfield, Astrophys. J.~675, 413--419 (2008)
\bibitem[Madsen et al  (2006)]{Madsen06} G.J. Madsen, R.J. Reynolds, and L.M. Haffner, Astrophys. J.~652, 401--425 (2006)
\bibitem[Minter and Balser (1997)]{Minter97a} A.H. Minter and D.S. Balser, Astrophys. J.~484, L133--L136 (1997)
\bibitem[Minter and Spangler (1997)]{Minter97b} A.H. Minter and S.R. Spangler, Astrophys. J.~485, 182--194 (1997)
\bibitem[Redfield and Linsky (2001)]{Redfield01} S. Redfield and J.L. Linsky, Astrophys. J.~551, 413--428 (2001)
\bibitem[Redfield and Linsky (2004)]{Redfield04} S. Redfield and J.L. Linsky, Astrophys. J.~613, 1004--1022 (2004)
\bibitem[Redfield and Linsky (2008)]{Redfield08} S. Redfield and J.L. Linsky, Astrophys. J.~673, 283--314 (2008)
\bibitem[Redfield (2009)]{Redfield09} S. Redfield, Space Sci. Rev.~143, 323--331 (2009)
\bibitem[Reynolds and Tufte (1995)]{Reynolds95} R.J. Reynolds and S.L. Tufte, Astrophys. J.~439, L17--L20 (1995)
\bibitem[Rickett (1990)]{Rickett90} B.J. Rickett, Annu. Rev. Astr. Ap.~28, 561--605 (1990)
\bibitem[Spangler (1991)]{Spangler91} S.R. Spangler, Astrophys. J.~376, 540--555 (1991)
\bibitem[Spangler (2007)]{Spangler07} S.R. Spangler, ``The Propagation Distance and Sources of Interstellar Turbulence'', in ``SINS-Small Ionized and Neutral Structures in the Diffuse Interstellar Medium'', M. Haverkorn and W.M. Goss, editors, Astr. Soc. Pac. Conference Series~367, 307--314 (2007)
\bibitem[Spangler et al (2010)]{Spangler10} S.R. Spangler, A.H. Savage, and S. Redfield, Nonlinear Proc. Geophys, in press, arXiv1008.1263 (2010)
\bibitem[Uzdensky (2007)]{Uzdensky07} D. Uzdensky, Astrophys. J.~671, 2139--2153 (2007)
\bibitem[Walker et al (2004)]{Walker04} M.A. Walker, D.B. Melrose, D.R. Stinebring, and C.M. Zhang, Mon. Not. Roy.Astro. Soc.~354, 43--53 (2004)
\bibitem[Welsh et al (2010)]{Welsh10} B.Y. Welsh, R. Lallement, J.L. Vergely, and S. Raimond, Astron. \& Astrophys..~510, A54, 1--23 (2010)
\end{thebibliography}
\end{document}